\documentclass[conference]{IEEEtran}
\IEEEoverridecommandlockouts

\usepackage{cite}
\usepackage{amsmath, amsthm, amssymb, amsfonts, bm, multirow}
\usepackage{algorithmic}
\usepackage{graphicx}
\usepackage{textcomp}
\usepackage{xcolor}
\usepackage{url}
\usepackage{caption, subcaption}
\usepackage[ruled,linesnumbered]{algorithm2e}
\usepackage{diagbox}
\usepackage{pgfplots}
\usepackage{graphicx}
\usepackage{booktabs}
\usepackage{cuted}
\newtheorem{assumption}{Assumption}
\def\BibTeX{{\rm B\kern-.05em{\sc i\kern-.025em b}\kern-.08em
    T\kern-.1667em\lower.7ex\hbox{E}\kern-.125emX}}
\begin{document}

\title{\Huge{Privacy-Aware Multi-Device Cooperative Edge Inference with Distributed Resource Bidding}}
\author{\IEEEauthorblockN{Wenhao Zhuang and Yuyi Mao}
\IEEEauthorblockA{Dept. of EEE, The Hong Kong Polytechnic University, Hung Hom,Hong Kong\\
Email: wenhao.zhuang@outlook.com, yuyi-eie.mao@polyu.edu.hk}
\vspace{-1em}
\thanks{
	This work was supported by the Start-Up Fund of the Hong Kong Polytechnic University under Project P0038174.
	}
}

\maketitle

\begin{abstract}
	Mobile edge computing (MEC) has empowered mobile devices (MDs) in supporting artificial intelligence (AI) applications through collaborative efforts with proximal MEC servers. Unfortunately, despite the great promise of device-edge cooperative AI inference, data privacy becomes an increasing concern. In this paper, we develop a privacy-aware multi-device cooperative edge inference system for classification tasks, which integrates a distributed bidding mechanism for the MEC server's computational resources. Intermediate feature compression is adopted as a principled approach to minimize data privacy leakage. To determine the bidding values and feature compression ratios in a distributed fashion, we formulate a decentralized partially observable Markov decision process (DEC-POMDP) model, for which, a multi-agent deep deterministic policy gradient (MADDPG)-based algorithm is developed. Simulation results demonstrate the effectiveness of the proposed algorithm in privacy-preserving cooperative edge inference. Specifically, given a sufficient level of data privacy protection, the proposed algorithm achieves $0.31$-$0.95\%$ improvements in classification accuracy compared to the approach being agnostic to the wireless channel conditions. The performance is further enhanced by $1.54$-$1.67\%$ by considering the difficulties of inference data.
\end{abstract}

\section{Introduction}

Driven by the tremendous success of deep neural networks (DNNs), edge artificial intelligence (AI) is envisaged to be a core service in the six-generation (6G) wireless networks~\cite{Letaief_JSAC_2022}. DNN-based inference is positioned as a critical task in edge AI systems, which analyzes data perceived by mobile devices (MDs) using DNN models~\cite{Mao_PIEEE_2024}. On one hand, on-device inference may not provide satisfactory accuracy due to resource restrictions. On the other hand, offloading data for inference at a mobile edge computing (MEC) server may lead to significant privacy risks. To mitigate these issues, device-edge cooperative inference has emerged~\cite{Li_TWC_2019}, which splits a DNN model into two parts for deployment at a mobile device and an MEC server, respectively. Intermediate features of data computed by the on-device model is transmitted to the MEC server. Accordingly, the server-based model derives the inference results.

Many efforts on device-edge cooperative inference have been made, aiming to strike a good balance between resource utilization and inference accuracy~\cite{Shao_MCOM_2020,Shao_JSAC_2022,Park_TCCN_2024}. Specifically, a three-step cooperative edge inference framework consisting of model split point selection, model compression, and feature encoding, was proposed in~\cite{Shao_MCOM_2020}, which optimizes the computation-communication tradeoff. The information bottleneck (IB) principle was first introduced in~\cite{Shao_JSAC_2022} to minimize the communication overhead while maximizing the inference accuracy of cooperative edge inference, based on which, a feature encoding scheme was developed to prune the redundant intermediate~feature dimensions. Moreover, the on-device and server-based models were jointly optimized for dynamic wireless channels and digital modulations in~\cite{Park_TCCN_2024}.

Although only the intermediate features are transmitted, local data are still vulnerable to privacy attacks. Model inversion attack~\cite{Hyunseok_2019,zecheng_attack_2021} is such a technique that allows adversaries to snoop local inference data from their intermediate features. To combat model inversion attacks in device-edge cooperative inference systems, several compute-efficient solutions have been developed~\cite{Ryu_WirelessNetw_2022,zecheng_attack_2021,Jiang_WCNC_2022,Wang_TWC_2024}. Differential privacy, which adds noise to intermediate features, was adopted in~\cite{Ryu_WirelessNetw_2022}. This technique was found particularly effective on image datasets with small intra-class variance. However, it was demonstrated in~\cite{zecheng_attack_2021} that additive noise may not provide sufficient privacy protection when a high inference accuracy is desired. Therefore, privacy-aware model partitioning was proposed in~\cite{zecheng_attack_2021,Jiang_WCNC_2022}. In addition, the IB principle was renovated with adversarial learning techniques in~\cite{Wang_TWC_2024} to end-to-end train the on-device and server-based models, which avoids transmitting features favored by model inversion attacks. 

Nevertheless, the scopes of most existing works on privacy-aware cooperative edge inference are limited to single-device systems. With multiple devices, the communication and computational resources need to be appropriately allocated while enforcing privacy protection mechanisms. This is extremely challenging when the MEC server is not fully trusted and decentralized decision making is necessary. In this paper, we develop a privacy-aware multi-device cooperative edge inference system, where MDs bid for the MEC server's services. Intermediate feature compression is adopted as the main approach to maximize classification accuracy while achieving satisfactory data privacy protection. Since it may be hard for the MDs to access the global system state, we formulate the design problem as a decentralized partially observable Markov decision process (DEC-POMDP) model~\cite{lowe2017multi}. A multi-agent deep deterministic policy gradient (MADDPG)-based algorithm is developed, where MDs determine the resource bidding values and intermediate feature compression ratios individually according to their uplink achievable rates, bidding budgets, and data classification difficulties.
Simulation results demonstrate that the proposed algorithm achieves competitive performance compared to a centralized upper bound. Moreover, it outperforms the baseline that is agnostic to the wireless channel conditions in classification accuracy by $0.31$-$0.95\%$. It also attains additional $1.54$-$1.67\%$ improvements by further considering the difficulties of inference data.

\begin{figure}
	\centering
	\includegraphics[width=0.75\linewidth]{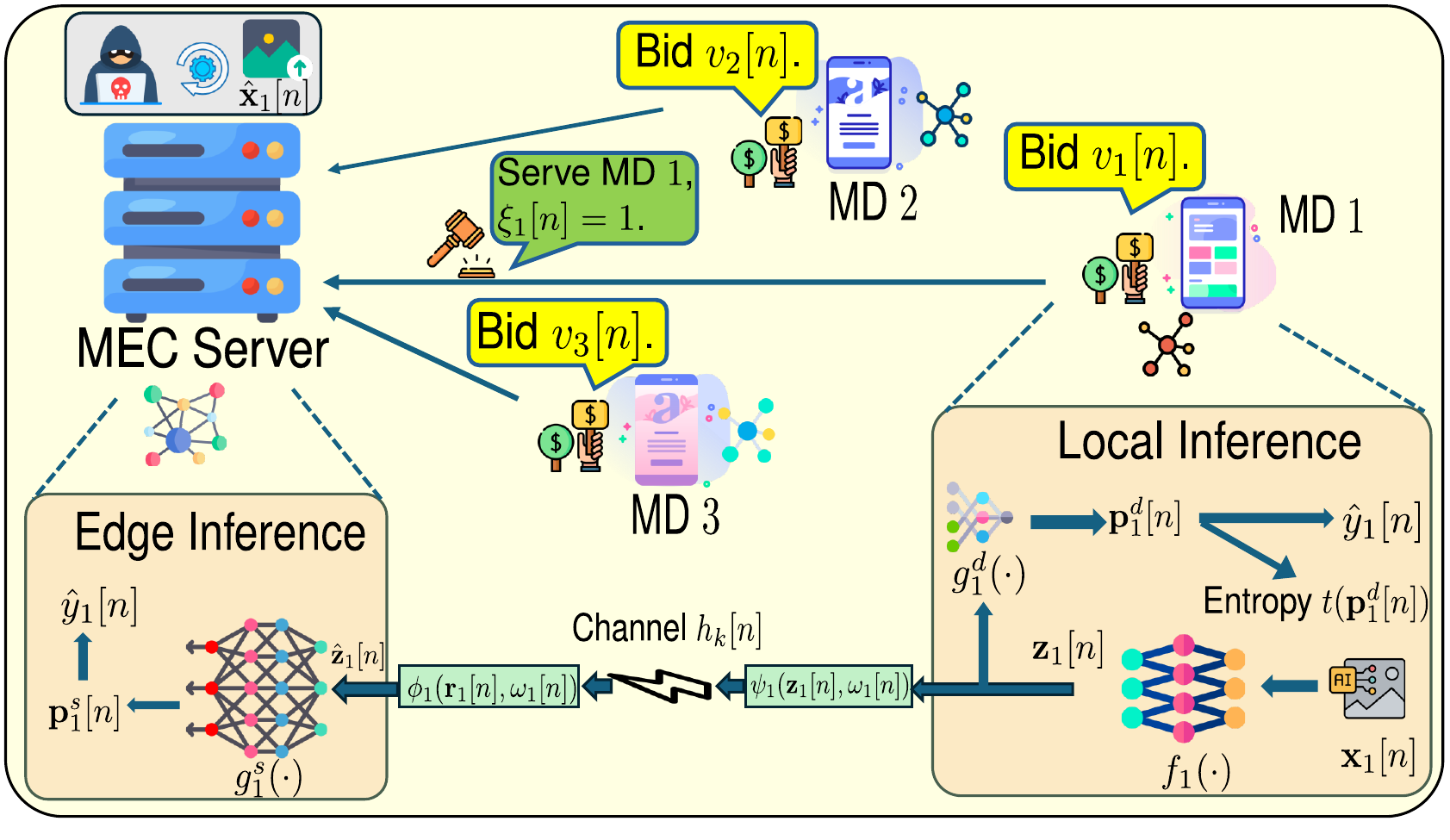}
	\caption{Privacy-aware multi-device cooperative edge inference.}
	\label{fig:system_model}
\end{figure}
	
\section{System Model}
We consider a multi-device cooperative edge inference system shown in Fig.~\ref{fig:system_model}, where $K$ MDs need to classify their local data using DNNs. Let $C_{k}$ be the number of classes to be considered at MD $k$. An MEC server is deployed to assist these MDs. However, the server is curious about their local data. The sets of MDs and time slots are denoted as $\mathcal{K} \triangleq \left\{1, \cdots, K\right\}$ and $\mathcal{N} \triangleq \left\{1, \cdots, N\right\}$, respectively. For simplicity, orthogonal channels are allocated to the MDs for uplink transmission. The maximum uplink throughput between MD $k$ and the MEC server at time slot $n$ is denoted as $c_{k}[n]$.
	
\vspace{-0.5em}
\subsection{Device-Edge Cooperative Inference}
\label{sec:device_edge_inference}
In the device-edge cooperative inference paradigm, the DNN model for classifying data of MD $k$ is split into two parts that are respectively deployed on device and the MEC server, namely the on-device model (denoted as $f_{k}(\cdot)$) and the server-based model (denoted as $g^{s}_{k}(\cdot)$). Given the local data $\mathbf{x}_{k}[n]$ of MD $k$ at time slot $n$, the $d_{k}$-dimensional intermediate feature computed by the on-device model is expressed as follows:
\begin{align}
	\mathbf{z}_{k}[n] = f_{k}(\mathbf{x}_{k}[n]),
\end{align}
of which, each dimension is encoded by $e_k$ bits. Then, $\mathbf{z}_{k}[n]$ is fed to a local classifier, denoted as $g^{d}_{k}(\cdot)$, to compute the logits (i.e., the probabilities of input data belonging to each class) as $\mathbf{p}^{d}_{k}[n] = \text{Softmax}\left(g^{d}_{k}(\hat{\mathbf{z}}_k[n])\right) \in [0,1]^{C_{k}}$.  

Let $\xi_{k}[n]\in\{0,1\}$ indicate whether MD $k$ is served by the MEC server at time slot $n$. When $\xi_{k}[n] = 0$, the logit of the local classifier is considered as the final classification result, i.e., $\mathbf{p}_{k}[n] = \mathbf{p}^{d}_{k}[n]$. When $\xi_{k}[n] = 1$, MD $k$ is served by the MEC server and the intermediate feature $\mathbf{z}_k[n]$ is first compressed by removing dimensions with the least magnitudes in order to protect local data from being disclosed. The compressed intermediate feature is expressed as follows:
\begin{align}
	\mathbf{r}_k[n] = \psi_{k} (\mathbf{z}_k[n], \omega_k[n]),
\end{align}
where $\psi_{k}(\cdot)$ denotes the compressor for MD $k$, $\omega_k[n] \in (0,1]$ is the compression ratio ($\omega_{k}[n]=1$ corresponds to the case without compression), and $\bm{\Omega}_k \triangleq \{\Omega_{k}[1],\cdots,\Omega_{k}[|\Omega_{k}|]\} $ gives the feasible set of $\omega_k[n]$. Hence, $\mathbf{r}_k [n]$ consists of $\omega_k[n] d_{k} e_{k}$ bits. Then, the compressed intermediate feature is transmitted to the MEC server and upsampled by zero interpolation. The upsampled intermediate feature is expressed as follows:
\begin{align}
	\hat{\mathbf{z}}_k[n] = \phi_{k}(\mathbf{r}_k[n], \omega_k[n]),
\end{align}
where $\phi_{k}(\cdot)$ denotes the upsampling operation at MD $k$. Afterwards, the server-based model takes $\hat{\mathbf{z}}_k[n]$ as input and obtains the logit as $\mathbf{p}_{k}[n] = \mathbf{p}^{s}_{k}[n] = \text{Softmax}\left(g^{s}_{k}(\hat{\mathbf{z}}_k[n])\right) \in [0,1]^{C_{k}}$. Finally, the class of $\mathbf{x}_{k}[n]$ is determined as follows:
\begin{equation}
	\hat{y}_{k}[n] = 
	\mathop{\arg \max}_{c\in\{1,2,\cdots,C_{k}\}} \left\{\mathbf{p}_{k,c}[n]\right\},
	\label{eq:edge_inference}
\end{equation}
where $\mathbf{p}_{k,c}[n]$ denotes the $c$-th dimension of $\mathbf{p}_{k}[n]$. Due to the limited on-device resources, $g^{d}_{k}(\cdot)$ is assumed to be less capable than $g^{s}_{k}(\cdot)$, i.e., its classification accuracy is lower. The determination of $\{\xi_{k}[n]\}$ is governed by the resource bidding outcomes, as detailed in the following sub-section.
\begin{figure}[t]
	\centering
	\begin{subfigure}{0.615\linewidth}
		\includegraphics[width=\linewidth]{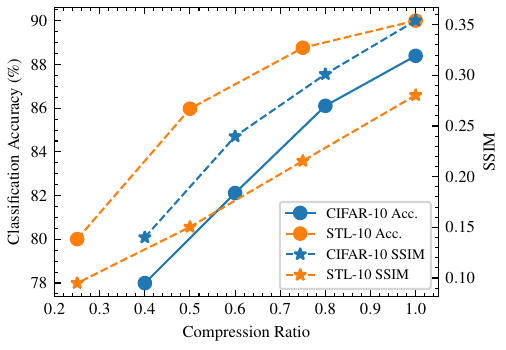}
		\label{fig:error}
	\end{subfigure}
	\begin{subfigure}{0.15\linewidth}
		\includegraphics[width=\linewidth]{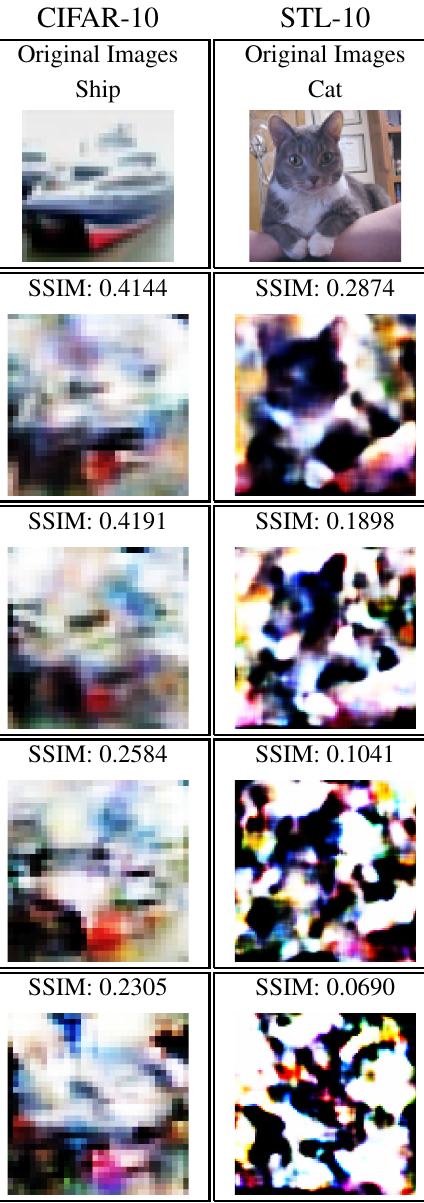}
		\label{fig:ssim}
	\end{subfigure}
	\vspace{-1em}
	\caption{Classification accuracy and SSIM value versus the compression ratio of intermediate features on the CIFAR-10 and STL-10 datasets. Details of the classification models and data reconstruction models are illustrated in Sec. V.}
	\label{fig:attack}
\end{figure}

\vspace{-0.5em}
\subsection{Distributed Resource Bidding}
\label{sec:resource_bidding}
We assume the MEC server can serve at most $U$ ($U \leq K$) MDs at each time slot. To facilitate rational allocation of the server's computational resources, a distributed resource bidding mechanism is implemented. Specifically, each MD is initialized with a resource bidding budget, denoted as $M_k^{b}$. At the beginning of time slot $n$, the MDs compete for the service of the MEC server by submitting their bids, denoted as $v_{k}[n] \geq 0$, which should not exceed the maximum value $v_{k}^{\max}$ and the remaining bidding budget. 

The MEC server admits $U$ MDs with the highest bids and provides them with cooperative inference services. Ties are broken randomly.
Besides, when a MD does not bid for the service, i.e., $v_{k}[n] = 0$, or the uplink achievable rate cannot support the successful delivery of the intermediate feature, i.e., $\omega_k[n] d_k e_k > c_k[n]$, it will not be served by the MEC server regardless the bidding value. 

\vspace{-0.5em}
\subsection{Model Inversion Attack and Privacy Leakage}
The MEC server is assumed to have access to the on-device model of $f_k(\cdot)$. As such, it uses a DNN model, denoted as $I_{\theta_k}(\cdot)$, to reconstruct the local data $\mathbf{x}_{k}[n]$ from $\mathbf{\hat{z}}_k[n]$. To train $I_{\theta_k}(\cdot)$, a small dataset $\mathcal{D}_k^{I}$ sharing similarities with the local data is also assumed available at the MEC server~\cite{zecheng_attack_2021}. This is valid when DNN models pre-trained on public datasets are used for the classification tasks~\cite{Yixiang_2024}. Specifically, $I_{\theta_k}(\cdot)$ is trained to minimize the $l_2$-loss between the reconstructed data and the original data as follows:
\begin{align}
	\mathcal{L}_{\text{inv}}(\theta_k) = \sum_{\mathbf{x}_k \in \mathcal{D}_k^{I}} \left\|I_{\theta_k}(f_k(\mathbf{x}_k)) - \mathbf{x}_k \right\|_2.
\end{align}
The reconstructed data at the MEC server is thus given as $\mathbf{\hat{x}}_k[n] = I_{\theta_k}(\mathbf{\hat{z}}_k[n])$.

To measure the reconstruction quality, i.e., the privacy leakage, a general distortion function between the reconstructed data $\mathbf{\hat{x}}$ and the original data $\mathbf{x}$, denoted as $\mathcal{L}_p(\mathbf{\hat{x}}, \mathbf{x})$, is introduced. For instance, the structural similarity (SSIM) index~\cite{Wang_2004} between $\hat{\mathbf{x}}$ and $\mathbf{x}$ defined below can be adopted for image data:
\begin{align}
	\mathcal{L}_p(\mathbf{\hat{x}}, \mathbf{x}) = \frac{(2\mu_{\mathbf{x}_{k}} \mu_{\mathbf{\hat{x}}} + C_1)(2\sigma_{\mathbf{x}, \mathbf{\hat{x}}} + C_2)}{(\mu^{2}_{\mathbf{x}} + \mu^{2}_{\mathbf{\hat{x}}} + C_1)(\sigma^{2}_{\mathbf{x}} + \sigma^{2}_{\mathbf{\hat{x}}} + C_2)},
\end{align}
where $2\mu_{\mathbf{x}}$ and $\sigma^{2}_{\mathbf{x}}$ denote the pixel sample mean and variance of $\mathbf{x}$, respectively. The covariance between $\mathbf{\hat{x}}$ and $\mathbf{x}$ is denoted as $\sigma_{\mathbf{x}, \mathbf{\hat{x}}}$ and $C_1, C_2$ are small positive empirical constants.\footnote{For RGB images, the SSIM index is obtained by averaging those of each color channel.}

In Fig.~\ref{fig:attack}, we evaluate the classification accuracy and~SSIM index by varying the compression ratio of intermediate features on the CIFAR-10 and STL-10 datasets, following the experimental setups detailed in Sec.~\ref{sec:simulation}. It is observed that both the accuracy and SSIM index increase with a higher compression ratio, which aligns with the observations in~\cite{zecheng_attack_2021}. Besides, the reconstructed images become indistinguishable from the originals when the SSIM index is below 0.26. 
In general, there exists a tradeoff between privacy leakage and classification accuracy, and the operating point can be adjusted by tuning the compression ratios of intermediate features. Moreover, the system performance should also be affected by the dynamic wireless channel capacities and resource bidding strategies at MDs. To maximize the classification accuracy of MDs while reducing data privacy leakage, we formulate a DEC-POMDP model in the next section. 

\vspace{-0.5em}
\section{Problem Formulation}
In this section, we first introduce an MDP model for the proposed system. Given the partially observable system state at MDs induced by distributed resource bidding, a DEC-POMDP model is further developed.

\vspace{-0.5em}
\subsection{MDP Model}
\label{sec:MDP}
We first make the assumption below on the inference data.
\begin{assumption}
	The inference data $\left\{\mathbf{x}_k[n]\right\}$ and the ground truth labels $\left\{y_k[n]\right\}$ are sampled from the joint distribution $P(\mathbf{X}_k, Y_k),\forall k\in\mathcal{K}$.
	\label{assumption:1}
\end{assumption}

The MDP model can be characterized by a four-tuple given as $(\mathcal{S}, \mathcal{A}, \mathcal{P}, \mathcal{R})$~\cite{Sutton_2018}. First, the state space $\mathcal{S}$ encompasses all possible states that provides necessary information for decision-making. The complete state of the proposed system includes the time slot index, inference data, uplink achievable rates, and remaining bidding budgets of all MDs. Secondly, the action space $\mathcal{A}$ includes all possible offloading decisions and feature compression ratios at MDs. Thirdly, the state transition function $\mathcal{P}$ represents the probability of transitioning to the next state given the current state and actions, which is determined by the joint distribution of inference data and uplink achievable rates, as well as the decision-making policy. Note that the expression of $\mathcal{P}$ is not available as the distributions $\{P\left(\mathbf{X}_{k},Y_{k}\right)\}$ cannot be known in advance. Finally, the reward function $\mathcal{R}$ evaluates the action in a given state, which should incorporate both classification accuracy and privacy leakage. 

The MDP model can be solved using deep reinforcement learning (DRL) algorithms. Specifically, a decision center, which has access to the complete system state, collects experiences in different episodes regarding the current state, actions of all MDs, state transitions, and perceived rewards over the $N$ time slots. With these experiences, the state transition function can be learned to optimize the offloading decisions of MDs and determine the feature compression ratios. In this centralized setup, distributed resource bidding is unnecessarily. Nevertheless, this solution is not applicable since the MDs individually bid for the MEC server's computational resources, resulting in a decentralized multi-agent system~\cite{DecPOMDPs_2016}. In addition, the system state is only partially observable at MDs, which further complicates algorithm designs. To tackle these challenges, we next develop a DEC-POMDP model.

\vspace{-0.5em}
\subsection{DEC-POMDP Model}
The DEC-POMDP model can be represented as $(\mathcal{S},\left\{\mathcal{A}_k\right\},$ $\mathcal{P}, \left\{\mathcal{O}_k\right\},  \left\{\mathcal{P}_k\right\}, \left\{\mathcal{R}_k\right\})$, where $\mathcal{S}$ and $\mathcal{P}$ follow the definitions in Sec.~\ref{sec:MDP} and other elements are detailed below.
\begin{itemize}
	\item \textbf{Action}: Following Sec.~\ref{sec:resource_bidding}, the action space of MD $k$, denoted as $\mathcal{A}_k$, involves all its feasible bidding values and feature compression ratios.
	We denote $\mathbf{A}_{k}[n] \triangleq [v_{k}[n], \omega_{k}[n]]$ as the action of MD $k$ at time slot $n$.

	\item \textbf{Observation}: At time slot $n$, each MD can only observe its uplink achievable rate, remaining bidding budget, and inference data. 
	Despite local models are available at MDs for classification, they generally tend to offload the processing of more challenging data to the MEC server to achieve higher accuracy. For effective decision making, it is essential to understand the difficulty of classifying the possibly high-dimensional data. 
	Because of this, we adopt the entropy of the logits obtained by the local model expressed in~\cite{Wu_2022_CVPR} as follows:
	\begin{align}
		{t}(\mathbf{p}^{d}_{k}[n]) = -\sum_{c=1}^{C_k} \mathbf{p}^{d}_{k, c}[n] \log(\mathbf{p}^{d}_{k, c}[n]),
	\end{align}
	which projects data to a positive scalar. According to~\cite{Wu_2022_CVPR}, data that are more challenging to be classified usually leads to a more uniform distribution of the logit, leading to higher values of ${t}(\mathbf{p}^{d}_{k}[n])$. Thus, the observation of MD $k$ at time slot $n$ is defined as follows:
	\begin{align}
		\mathcal{O}_{k}[n] \triangleq [n, c_{k}[n], M_{k}[n], {t}(\mathbf{p}^{d}_{k}[n])],
	\end{align}
	where $M_k[n]$ denotes the remaining bidding budget with $M_{k}[1] = M_k$ and $M_{k}[n+1] = M_{k}[n] - \xi_{k}[n] v_{k}[n]$.
	 \item \textbf{Observation Probability}: We denote the probability of MD $k$ observing $\mathcal{O}_{k}[n+1]$ when actions $\{\mathbf{A}_{k^{\prime}}[n]\}$ are executed in state $\mathcal{S}[n]$ as $\mathcal{P}_{k}(\mathcal{O}_{k}[n+1] | \mathcal{S}[n], \{\mathbf{A}_{k^{\prime}}[n]\})$. 
	Note that the observation probability is even more arduous to learn compared to the state transition function as the decision-making policies of other MDs, i.e., $\{\mathbf{A}_{k^{\prime}}[n]\} \setminus \mathbf{A}_{k}[n]$, are not available at MD $k$. 

	\item \textbf{Reward}: The rewards of the MDs are individually evaluated and designed to maximize the classification accuracy while ensuring data privacy as follows: 
	\begin{align}
		\mathcal{R}_k (\mathbf{A}_k[n] | \mathcal{S}[n]) & \triangleq t_{1, k} \cdot \mathcal{L}_{a}(\mathbf{p}_{k}[n], \hat{y}_{k, n}) \nonumber \\
		& \ \ \ - t_2 \cdot \mathcal{L}_{p}(\hat{\mathbf{x}}_k[n], \mathbf{x}_k[n]),
		\label{eq:reward}
	\end{align}
	where $\{t_{1, k}\}, t_2$ are positive weight coefficients and $\mathcal{L}_{c}(\mathbf{p}_{k}[n], {y}_{k, n}) \triangleq -\sum_{c=1}^{C_k} \mathbf{p}_{k, c}[n] \log({y}_{k, n})$ is the cross-entropy loss function for evaluating the classification accuracy.
\end{itemize}
Under the DEC-POMDP model, each MD aims to find a policy that generates an action $\mathbf{A}_k[n]$ based on its observation $\mathcal{O}_{k}[n]$ to maximize the expected cumulative reward $\mathbb{E}[\sum_{n=1}^{N} $ $ \mathcal{R}_k ( \mathbf{A}_k[n] | \mathcal{S}[n])]$. Next, we propose an \textbf{MADDPG}-based algorithm to distributedly learn effective policies for the MDs.

\section{Proposed Solution}
\label{sec:algorithm}
In this section, we develop an \textbf{MADDPG}-based algorithm, where the learning process includes the experience collection and model update processes. The proposed algorithm can learn the distributions of inference data and uplink achievable rates, as well as the intricate relationship between the feature compression ratios, classification accuracies, and privacy leakage for policy optimization, i.e., the bidding value and feature compression ratio of each MD.

\vspace{-0.5em}
\subsection{Critic-Actor Networks}
In the proposed algorithm, critic-actor networks are employed for decision-making policy learning. The actor network $\pi_{k}$ is employed to generate the action of MD $k$ based on its local observation. The action of each MD contains the continuous bidding value and the discrete selection of feature compression ratio. To this end, we construct the actor network $\pi_{k}$ with two sub-networks, namely $\pi_{v, k}(\mathcal{O}_k[n]; \bm{\xi}_{k})$ and $\pi_{\omega, k}(\mathcal{O}_k[n]; \bm{\lambda}_{k})$ for generating the bidding values and the logits of feature compression ratio selection, respectively. Hence, the action of MD $k$ at time slot $n$ is determined as $\pi_k(\mathcal{O}_k[n] ; \bm{\Theta}_k) \triangleq \left[\pi_{v, k}(\mathcal{O}_k[n]; \bm{\xi}_{k}), \Omega_k[i]\right]$ with $\bm{\Theta}_{k} \! \triangleq  \{\bm{\xi}_{k},$ $ \bm{\lambda}_{k}\}$, $i =\arg\max\{\pi_{\omega, k, i}(\mathcal{O}_k[n]; \bm{\lambda}_{k})\}$ and $\pi_{\omega, k, i}(\mathcal{O}_k[n]; \bm{\lambda}_{k})$ denoting the $i$-th dimension of $\pi_{\omega, k}(\mathcal{O}_k[n]; \bm{\lambda}_{k})$. Meanwhile, a critic network, denoted as $Q(\{\mathcal{O}_{k^{\prime}}[n], \mathbf{A}_{k^{\prime}}[n]\}_{k^{\prime}}; \bm{\Phi}_k)$, at MD $k$ is used to approximate the state-action function, i.e., the expected cumulative reward of MD $k$ starting from time slot $n$ given the joint actions and observations of all MDs, where $\bm{\Phi}_k$ are trainable parameters. 

To mitigate the potential overestimation of the expected cumulative reward by the critic networks during training, we respectively introduce target-critic networks and target-actor networks, parameterized by $\bm{\Phi}^{\prime}_k$ and $\bm{\Theta}^{\prime}_k$, which share the same architecture respectively as the critic and actor networks, to help stabilize the model update process~\cite{lowe2017multi}.

\begin{algorithm}[t]
	\footnotesize
	\caption{Proposed {MADDPG}-based Algorithm.}
	\label{alg:algorithm}
	\KwIn{Initialize $M$, $N$, $\sigma_z^2$, $\tau_{Q}, \tau_{\pi}$, $\alpha_1$, $\alpha_2$, and $t^{\max}$.}
	\KwOut{$\left\{\bm{\Phi}^{\prime}_k\right\}$ and $\{\bm{\Theta}^{\prime}_k\}$} 
	\For{$t=1$ to $t^{\max}$}{
		Set $M_k[1] = M_k$, for all $k$\;
		\textbf{// Experience Collection Process} \\
		\For{$n = 1$ to $N$}{
			Randomly sample $c_k[n]$, $(\mathbf{x}_k[n],y_k[n])$ for all $k$ \;
			\For{each MD $k$}{
				Determine $\hat{v}_{k, n}$ and $\hat{\omega}_{k}[n]$ with~\eqref{eq:noise},~\eqref{eq:noise1} \;
			}
		Update $M_k[n+1]$ for all $k$ \;
		}
		Store experience $\mathbf{e}_t$ in the $\mathcal{D}$\;
		\textbf{// Model Update Process} \\
		Sample $\mathcal{E}$ from $\mathcal{D}$ and update $\bm{\Phi}_k$, $\bm{\Theta}_k$ with \eqref{eq:phi_loss},~\eqref{eq:update_theta} \;
		Soft-update $\bm{\Phi}^{\prime}_k$ and $\bm{\Theta}^{\prime}_k$ \;
	}
\end{algorithm}

\vspace{-0.5em}
\subsection{Experience Collection}
In the \textbf{MADDPG}-based algorithm, the current state, action, next state, and reward of each MD are collected as experience and used for updating the critic-actor networks. During experience collection process, the MDs are encouraged to explore their action spaces via the exploitation-exploration strategy. On one hand, the bidding values generated by the actor networks are perturbed by introducing random factors as follows:
\begin{align}
	\tilde{v}_{k}[n] = \pi_{v, k}(\mathcal{O}_k[n]; \bm{\xi}_{k}) + z_{k}[n],
	\label{eq:noise}
\end{align}
where $z_{k}[n] \sim \mathcal{N}(0, \sigma_z^{2})$ denotes the Gaussian noise with zero mean and variance $\sigma_z^{2}$. Then, the bidding value is clipped to the feasible range, i.e., $\hat{v}_{k}[n] = \max\left\{0, \min\left\{\tilde{v}_{k}[n], M_k[n]\right\}\right\}$.
On the other hand, simply selecting the feature compression ratio with the highest probability in $\pi_{\omega, k}(\mathcal{O}_k[n]; \bm{\lambda}_{k})$ is deterministic and does not facilitate exploration. Hence, we sample the feature compression ratio from the discrete distribution $\pi_{\omega, k}(\mathcal{O}_k[n]; \bm{\lambda}_{k})$. However, since the sampling process is non-differentiable, the back-propagation algorithm cannot be used for updating $\{\bm{\Theta}_k\}$. Hence, the Gumbel-Softmax trick~\cite{jang2016categorical} is adopted to approximate the sampling process as follows:
\begin{align}
	\hat{\pi}_{\omega, k}(\mathcal{O}_k[n]; \bm{\lambda}_{k}) = \text{Softmax}\left(\frac{\log(\pi_{\omega, k}(\mathcal{O}_k[n]; \bm{\lambda}_{k}))  + g} {\tau_{\omega}}\right), 
	\label{eq:noise1}
\end{align}
where $\log(\cdot)$ returns the element-wise logarithm of the input, $g$ is a Gumbel random variable, and $\tau_{\omega}$ is a control parameter for the exploration process. Specifically, the sampling policy becomes more deterministic as $\tau_{\omega} \rightarrow 0$ and more random as $\tau_{\omega} \rightarrow \infty$. Note that $\hat{\pi}_{k}(\mathcal{O}_k[n]; \bm{\Theta}_{k})$ is differentiable with respect to $\bm{\Theta}_{k}$. The action of MD $k$ during the experience collection process is denoted as $\mathbf{A}_k[n] = [\hat{v}_{k}[n], \Omega_k[i]]$ with $i = \arg\max\{\hat{\pi}_{\omega, k, i}(\mathcal{O}_k[n]; \bm{\lambda}_{k})\}$ and $\hat{\pi}_{\omega, k, i}(\mathcal{O}_k[n]; \bm{\lambda}_{k})$ denoting the $i$-th dimension of $\hat{\pi}_{\omega, k}(\mathcal{O}_k[n]; \bm{\lambda}_{k})$.

In the proposed DEC-POMDP model, a policy is evaluated over entire episodes, i.e., the cumulative reward over $N$ time slots. This contrasts with conventional DRL approaches that maximize the long-term rewards. Thus, a hierarchical replay buffer~\cite{xiaowen_2024} $\mathcal{D}$ with capacity $M$ working in a first-in-first-out manner is adopted. Specifically, we denote $\mathbf{d}_{k}[n] \triangleq (\mathcal{O}_k[n],$ $ \mathbf{A}_k[n], \mathcal{O}_k [n+1], \mathcal{R}_k(\mathbf{A}_{k}[n] | \mathcal{S}[n]))$ as the experience of MD $k$ collected at time slot $n$. Then, a piece of episodic experience is constructed as $\mathbf{e}_t \triangleq \left\{\mathbf{d}_{k}[n]\right\}$ and stored in $\mathcal{D}$.

\vspace{-0.5em}
\subsection{Model Update}
\label{sec:model_update}
Denote $\alpha_{1}$ and $\alpha_{2}$ as the learning rates for updating the critic and actor networks, respectively. During the model update process, a batch of episodic experience $\mathcal{E}$ is sampled from $\mathcal{D}$ and the critic networks are updated by minimizing the estimated error of the state-action value function according to $\bm{\Phi}_k \leftarrow \bm{\Phi}_k - \alpha_{1} \nabla_{\Phi_k} \mathcal{L}(\bm{\Phi}_k)$ and 
\vspace{-0.5em}
\begin{align}
	\mathcal{L}(\bm{\Phi}_k) &\! = \! \frac{1}{|\mathcal{E}|} \sum_{\mathbf{d}_k[n]\in \mathcal{E}} \sum_{n=1}^{N} \left( z_k[n] \! - \! Q(\{\mathcal{O}_{k^{\prime}}[n], \mathbf{A}_{k^{\prime}}[n] \}_{k^{\prime}}; \bm{\Phi}_k) \right)^2, \label{eq:phi_loss}
\end{align}
where $z_k[n] \!\triangleq \! Q( \{ \mathcal{O}_{k^{\prime}}[n+1], \pi_{k^{\prime}}(\mathcal{O}_{k^{\prime}}[n+1];$ $\bm{\Theta}^{\prime}_{k^{\prime}}) \}_{k^{\prime}} ; \bm{\Phi}^{\prime}_k) +  \mathcal{R}_k (\mathbf{A}_k[n]| \mathcal{S}[n])$. The actor network is updated by maximizing the output of the critic network as follows:
\vspace{-0.5em}
\begin{align}
	\bm{\Theta}_k  & \leftarrow \bm{\Theta}_k + \frac{\alpha_{2} }{|\mathcal{E}|} \sum_{d_k[n] \in \mathcal{E}} \sum_{n=1}^{N} \nabla_{\bm{\Theta}_k} \pi_k(\mathcal{O}_k[n]; \bm{\Theta}_k) \nonumber \\
	& \ \ \ \quad \quad \quad \cdot  \nabla_{\mathbf{A}_k[n]} Q\left( \left\{\mathcal{O}_{k^{\prime}}[n], \mathbf{A}_{k^{\prime}}[n]\right\}_{k^{\prime}}; \bm{\Phi}_k  \right), 
	\label{eq:update_theta}
	\end{align}
Then, target-critic network and target-actor networks are soft-updated to mitigate the estimation bias as $\bm{\Phi}^{\prime}_k \leftarrow \tau_{Q} \bm{\Phi}_k + (1-\tau_{Q}) \bm{\Phi}^{\prime}_k$ and $\bm{\Theta}^{\prime}_k \leftarrow \tau_{\pi} \bm{\Theta}_k + (1-\tau_{\pi}) \bm{\Theta}^{\prime}_k$ with $\tau_{Q}, \tau_{\pi} \in (0, 1]$ being the soft-update hyperparameters.

The proposed \textbf{MADDPG}-based algorithm is summarized in Algorithm~\ref{alg:algorithm}: First, the system state are initialized at the beginning of each episode (line 2). Then, the MDs execute their policies based on the proposed exploitation-exploration strategy and the experiences are stored in the replay buffers (lines 3-9). To effectively learn the policies, a batch of experiences $\mathcal{E}$ are sampled from the replay buffer and the critic-actor networks are updated (lines 10-12). The learning process is repeated until the maximum episode $t^{\max}$ is reached.

\vspace{-0.5em}
\section{Simulation Results}
\label{sec:simulation}
We evaluate the proposed \textbf{MADDPG}-based algorithm in a cooperative edge inference system with $K=2$ MDs performing image classification tasks on the STL-10 and CIFAR-10 datasets, respectively. The MEC server can serve at most $U=1$ MD in each time slot. The DNN model architectures are given in Table~\ref{tab:parameters} for classifying the STL-10 and CIFAR-10 data. The data reconstruction models are constructed at the MEC server by inverting the architecture of the corresponding on-device model $f_{k}\left(\cdot\right)$.The datasets are divided into three parts for different purposes: 1) training the classification and data reconstruction models; 2) training the \textbf{MADDPG}-based algorithm; and 3) testing, which respectively contain $5,000$, $6,000$ and $1,000$ ($50,000$, $6,000$ and $1,000$) samples for the STL-10 (CIFAR-10) dataset.
Other simulation parameters are set as follows: $N=10$, $M_1=M_2=5$, $M=2,000$, $t^{\max}=5,000$, $\sigma_z^2=0.1$, $\tau_{Q}=0.01$, $\tau_{\pi}=0.01$,  $|\mathcal{E}|=256$, $\alpha_1 \! = \! \alpha_2 \! = \! 10^{-3}$, and $c_k[n] = \delta B \log_2 (1 + \frac{ p_k |h_k[n]|^2}{B N_0})$ kbits per time slot with $\delta=0.1$ s, $B=50$ kHz, $N_0=79$ dBm and $h_k[n] \sim \mathcal{CN}(0, 1)$. The sets of admissible compression ratios for MD $1$ and MD $2$ are given as $\bm{\Omega}_1 \triangleq \left\{1.0, 0.8, 0.6, 0.4\right\}$ and $\bm{\Omega}_2 \triangleq \left\{1.0, 0.75,  0.5, 0.25\right\}$, respectively. All critic-actor network architectures are configured as three-layer DNNs with $32$ neurons in each hidden layer. For comparisons, we consider the following baselines:
\begin{itemize}
	\item \textbf{DQN}: The MEC server employs a deep Q-network (DQN) to select a subset of MDs to serve and determine their feature compression ratios. The DQN is trained assuming the uplink achievable rates and the entropy of logits $t(\mathbf{p}^{d}_{k}[n])$ are available at the MEC server.
	\item \textbf{SIB}: In the statistical information-based (SIB) algorithm, each MD selects the minimum compression ratio that achieves a given SSIM index value on the average sense. The MEC server randomly selects a subset of MDs to serve in each time slot regardless of their uplink achievable rates.
\end{itemize}
Three variants of the proposed \textbf{MADDPG}-based algorithm, dubbed \textbf{MADDPG-DD}, \textbf{MADDPG-DT}, and \textbf{MADDPG-MC}, are also investigated. Specifically,~\textbf{MADDPG-DD} disregards the data classification difficulties by removing $t(\mathbf{p}^{d}_{k}[n])$ from the observation of MDs. \textbf{MADDPG-DT} directly transmits the intermediate features of MDs to the MEC server without compression, while \textbf{MADDPG-MC} compresses the intermediate feature with the maximum value in $\Omega_k$ for each~MD.

\begin{table}[t]
\centering
\footnotesize
\setlength{\tabcolsep}{3pt}
\renewcommand{\arraystretch}{1.2}
\begin{tabular}{c|c|c|c}
\hline
\multicolumn{1}{c|}{Model} &  
\multicolumn{1}{c|}{$f_{k}(\cdot)$} &
\multicolumn{1}{c|}{$g^{d}_{k}(\cdot)$} &
\multicolumn{1}{c}{$g^{s}_{k}(\cdot)$}\\
\hline
\begin{tabular}[c]{@{}c@{}}MD $1$\end{tabular} & 
\begin{tabular}[c]{@{}c@{}}
	$[\text{VGG Block}]\times 2$ \\
	Maxpool \\
\end{tabular} & 
\begin{tabular}[c]{@{}c@{}}
	FC(8192, 4096)\\
	FC(4096, 4096)\\
	FC(4096, 10)\\
\end{tabular} & 
\begin{tabular}[c]{@{}c@{}}
	$[\text{VGG Block}] \times 11$ \\
	Maxpool \\
	FC(512, 4096)\\
	FC(4096, 4096)\\
	FC(4096, 10)\\
\end{tabular} 
\\
\hline
\begin{tabular}[c]{@{}c@{}}MD $2$\end{tabular} &
\begin{tabular}[c]{@{}c@{}}
	Conv(7, 64, 2) \\
	Maxpool \\
	$[\text{ResNet Block}]\times 3$ \\
	Maxpool \\
\end{tabular} & 
\begin{tabular}[c]{@{}c@{}}
	FC(128, 10)\\
\end{tabular} &
\begin{tabular}[c]{@{}c@{}}
	$[\text{ResNet Block}] \times 13$ \\
	Maxpool \\
	FC(512, 10)\\
\end{tabular} \\
\hline
\end{tabular}
\caption{DNN model architectures for classifying the STL-10 and CIFAR-10 data on MD $1$ and MD $2$, respectively, where `Conv$(k, m, n)$' denotes a convolutional layer with kernel size $k$, output channels $m$, and stride $n$, and `FC$(a, b)$' denotes a fully connected layer with input size $a$ and output size $b$. The ReLU activation function is omitted in the table.}
\label{tab:parameters}
\end{table}
\begin{figure}[t]
	\vspace{-1em}
	\centering
	\begin{subfigure}{0.42\linewidth}
		\includegraphics[width=\linewidth]{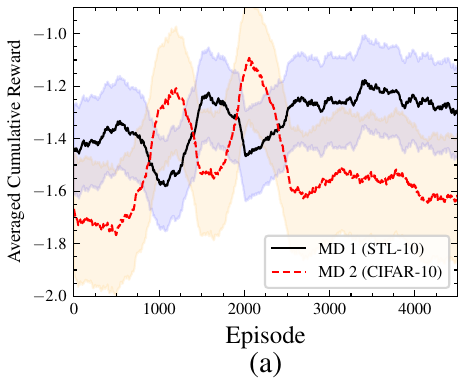}
	\end{subfigure}
	\begin{subfigure}{0.42\linewidth}
		\includegraphics[width=\linewidth]{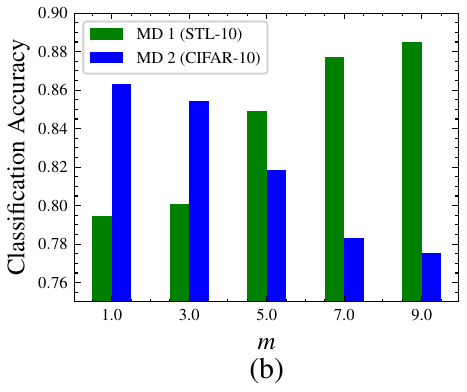}
	\end{subfigure}
	\vspace{-0.5em}
	\caption{(a) Averaged cumulative reward achieved by the proposed \textbf{MADDPG}-based algorithm over $5,000$ episodes during the learning process. The weight coefficients of~\eqref{eq:reward} are set as $t_{1, 1}=t_{1, 2}=0.2, t_2=0.8$. (b) Classification accuracy versus the value of $m$.}
	\label{fig:rewards}
\end{figure}

The rewards achieved by the MDs during the learning process of \textbf{MADDPG} are depicted in Fig.~\ref{fig:rewards}(a), which gradually improve over episodes, indicating that effective policies for the MDs are learned. Meanwhile, the results also demonstrate the competition between the two MDs, e.g., the reward of MD $1$ increases while that of MD $2$ decreases between Episodes $500$-$1,000$ and $1,750$-$2,000$. In Fig.~\ref{fig:rewards}(b), we configure the budgets of the MDs as $M_1=m$ and $M_2=10-m$, where $m \in \{1, 3, 5, 7, 9\}$ and $t_{1, 1} = t_{1, 2}=0.9$, $t_{2}=0.1$. The results show an inverse relationship between the classification accuracies of MD $1$ and MD $2$, as the MD with a higher budget gains more advantage in bidding for the MEC server's cooperative inference service on testing datasets.

\begin{figure}[t]
	\centering
	\begin{subfigure}{0.42\linewidth}
		\includegraphics[width=\linewidth]{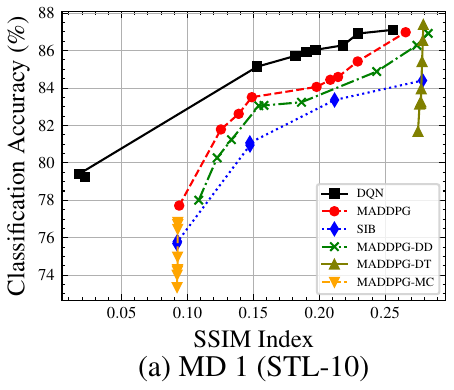}
	\end{subfigure}
	\begin{subfigure}{0.42\linewidth}
		\includegraphics[width=\linewidth]{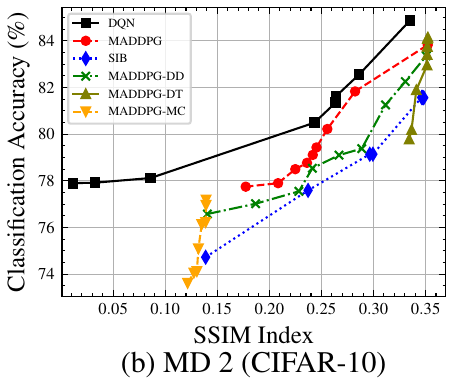}
	\end{subfigure}
	\vspace{-0.5em}
	\caption{Classification accuracy versus SSIM index value with testing datasets.	The classification accuracy of local inference and edge inference without feature compression are 73.0\% and 90.0\% for MD 1, and 72.0\% and 88.4\% for MD 2, respectively.}
	\label{fig:tradeoffs}
\end{figure}

In Fig.~\ref{fig:tradeoffs}, we compare the performance of different algorithms in terms of the classification accuracy and SSIM index value by varying $\{t_{1, k}\}$ and $t_2$. Since \textbf{DQN} has access to the global system information and learn a centralized decision-making policy, it achieves the best performance, i.e., the highest classification accuracy for a given SSIM index value. Despite each MD can only partially observe the system state, \textbf{MADDPG} still attains competitive performance. 
For example, if we target an SSIM index value of $0.26$ on both MDs (which implies a satisfactory level of privacy protection as shown in Fig.~\ref{fig:attack}), the classification accuracy of \textbf{MADDPG} only degrades by $1.32\%$ for MD $1$ and $0.27\%$ for MD $2$ compared to \textbf{DQN}. Nevertheless, \textbf{MADDPG-DD} suffers from a classification accuracy degradation by $1.54\%$ for MD $1$ and $1.67\%$ for MD $2$ compared with \textbf{MADDPG}. This is because $t(\mathbf{p}_{k}^{d}[n])$ provides valuable information on the local classification results, so the limited bidding budget is better utilized by offloading more challenging data to the MEC server with an appropriate feature compression ratio can be adjusted. Besides, the performance of \textbf{SIB} further degrades by $0.31$-$0.95\%$ compared to \textbf{MADDPG-DD} due to its agnostic nature to dynamic wireless channel conditions. Moreover, the classification accuracy of \textbf{MADDPG} under a high SSIM index value approaches that achieved by \textbf{MADDPG-DT}, implying that the proposed algorithm tends not to compress the intermediate features for accuracy when data privacy leakage is not a concern. In addition, the SSIM index values achieved by \textbf{MADDPG-DT} and \textbf{MADDPG-MC} are rather stable as the feature compression ratios are constants, yet they fail to properly balance the classification accuracy and data privacy~leakage. 

\section{Conclusion}
In this paper, we investigated a privacy-aware multi-device cooperative inference system with distributed resource bidding, where intermediate feature compression is adopted as the main principle for reducing data privacy leakage. We casted the distributed optimization of bidding values and feature compression ratios under the decentralized partially observable Markov decision process model and developed an \textbf{MADDPG}-based algorithm. Simulation results showed that the proposed algorithm achieves competitive performance compared to a centralized upper bound and validated the importance of the joint consideration on data classification difficulties and feature compression. Our future work will focus on further optimizing the computational resource allocation at the MEC server.

\vspace{-0.5em}
\bibliographystyle{IEEEtran}
\bibliography{ref}
\end{document}